\documentclass[11pt]{article}
\usepackage{amsmath}
\usepackage{amssymb}
 \allowdisplaybreaks[4]
\makeatother       % '@' is restored as a "non-letter" character for TeX

\makeatletter      % '@' is now a normail "letter" for TeX
\@addtoreset{equation}{section}
\makeatother       % '@' is restored as a "non-letter" character for TeX

\title{ \Huge The determinant representation of the gauge transformation for the discrete KP hierarchy }
\author{  Liu Shaowei\dag,
Cheng Yi\dag, He Jingsong\ddag\thanks{Corresponding author, Email:hejingsong@nbu.edu.cn,jshe@ustc.edu.cn},\\
\dag\scriptsize Department of  Mathematics, University of Science
and
Technology of China, Hefei, 230026 Anhui,P.R. China \\
\ddag\scriptsize Department of  Mathematics, Ningbo University,
Ningbo, 315211 Zhejiang, P.R. China
 }
 \vspace{1cm}
\date{}
\begin{document}
\maketitle
\begin{center}
\begin{center}{\bf Abstract}\end{center}
\vspace{1cm}
\begin{minipage}{12cm}
 {A successive gauge transformation operator $T_{n+k}$ for the discrete KP(dKP) hierarchy is defined, which is involved with two
types of gauge transformations operators. The determinant representation of the $T_{n+k}$ is established,and then it is used to
get a new $tau$ function $\tau^{(n+k)}_\triangle$ of the dKP  hierarchy  from an initial $\tau_\triangle$. In this process, we
introduce a generalized discrete Wronskian determinant and some useful properties of discrete difference operator.}
 \end{minipage}
\end{center}
{\bf 2000 MR Subject Classification} \ \  35Q51, 37K10 \\

\vspace{2cm}
\newpage
%\begin{center}
\section*{\S 1.\ \ Introduction}
%\end{center}
\setcounter{section}{1}

The discrete KP(dKP) hierarchy\cite{oe,hi,dk,ku} is an interesting  object in the research of the discrete integrable systems and the
discretization of the integrable systems\cite{abs,ab,bs2,su1}. Naturally, there exist some similar properties between dKP and KP
hierarchy\cite{jim}, such as $tau$ function\cite{hi,jim,dk2}, Hamiltonian structures\cite{dk,ku} and gauge transformation\cite{oe,ch,hlc},
etc. In particular, Gauge transformation is one kind of effective way to construct the solution of the  integrable systems, both
continuous and discrete. Since Chau et al.\cite{ch} have introduced two types of  gauge transformation operators for the KP hierarchy,
the similar operator for  the constrained KP hierarchy\cite{oe2,chaw2,ar,hlc2}, q-KP hierarchy \cite{tu,tu2,hlc3} and discrete KP hierarchy
\cite{oe} have also been given. Oevel \cite{oe} has given explicitly three types of gauge transforation operators of the dKP hierarchy,
which are called DT, adjoint DT and binary DT. However, DT and adjoint DT are two elementary gauge transformation operators, because a
binary DT consists of a DT and an adjoint DT. Here DT and adjoint DT of the dKP hierarchy  are regarded as  discrete analogues of the
$T_D$ and $T_I$ \cite{ch} of the KP hierarchy, so we shall denote it by $T_d$ and $T_i$ respectively in following sections. Furthermore,
in theorem 3 of \cite{oe} Oevel has also considered n-fold iteration of $T_d$ and binary DT for the dKP hierarchy \cite{oe}.\footnote{The expression of dKP and n-fold iteration of
gauge transformation in \cite{oe} is different from here, but the  essence is the same.} On the other hand, the determinant representation
\cite{hlc} of the gauge transformation operators provides a simple method to construct the transformed $\tau$ function \cite{hlc2,hlc3,
hcr, hwc} for several special cases of the KP hierarchy and q-KP hierarchy. In this paper we shall extend these results to discrete KP hierarchy.
We combine the two elementary types of gauge transformation operators ($T_d$ and $T_i$) for the dKP hierarchy and let them act n times and k times respectively.
So we get the combined gauge operator $T_{n+k}$ and its the determinant representation, then use it to construct new tau function
$\tau^{(n+k)}_\triangle$ of the dKP hierarchy from an initial tau function.

The organization of the paper is as follows. In section 2, we give a
brief description of the discrete KP hierarchy and prove some useful
properties of the discrete operators. In section 3, based on the
Oevel's two types elementary  gauge transforation operator\cite{oe},$T_d$ and $T_i$,
 we give the determinant representation of the gauge transformation operator
$T_{n+k}$. In section 4, we  shall construct the
$\tau_\triangle^{(n+k)}$ function of the dKP after the successive
gauge transformations by using previous results. Section 5 is
devoted to the conclusions and discussions.

%%%%%%%%%%%%%%%%%%%%%%%%%%%%%%%%%%%%%%%%%%%%%%%%%%%%%%%%%%%%%%%%%%%%%%%%%%%%%%%%%%%%%%%%%%%%%%%%%%%%%%%%%%%%%%%%%
%%%%%%%%%%%%%%%%%%%%%%%%%%%%%%%%%%%%%%%%%%%%%%%%%%%%%%%%%%%%%%%%%%%%%%%%%%%%%%%%%%%%%%%%%%%%%%%%%%%%%%%%%%%%%%%%%
%\begin{center}
\section*{\S 2.  The discrete KP and operators }
%\end{center}
\setcounter{section}{2}

        To be self-contained, we give a brief introduction to dKP hierarchy
based on detailed research in \cite{hi}. Here we denote by $\Gamma$
and $\triangle$ respectively, the shift and the difference operators
acting on the associative ring F of functions. Where
\[F=\left\{ f(n)=f(n,t_1,t_2,\cdots,t_j,\cdots);  n\in\mathbb{Z}, t_i\in\mathbb{R}
\right\},
\]
and
\[\Gamma f(n)=f(n+1),
\]
\[\triangle f(n)=f(n+1)-f(n)=(\Gamma -I)f(n).
\]
Where $I$ is the identity operator. Define the following operation,
for any$j\in\mathbb{Z}$
\begin{equation}\triangle^j\circ
f=\sum^{\infty}_{i=0}\binom{j}{i}(\triangle^i
f)(n+j-i)\triangle^{j-i},\hspace{.3cm}
\binom{j}{i}=\frac{j(j-1)\cdots(j-i+1)}{i!}.\label{81}
\end{equation}
So we obtain an associative ring $F(\triangle)$ of formal pseudo
difference operators, with the operation $``+"$ and $``\circ"$
\[F(\triangle)=\left\{R=\sum_{j=-\infty}^d f_j(n)\triangle^j, f_j(n)\in
R, n\in\mathbb{Z}\right\}.
\]
We also denote by $R_+=\sum_{j=0}^d f_j(n)\triangle^j$, the positive
difference part of $R$ and by $R_-=\sum_{j=-\infty}^{-1}
f_j(n)\triangle^j$, the Volterra part of $R$. Also define the
adjoint operator to the $\triangle$ operator by $\triangle^*$,
\[\triangle^* f(n)=(\Gamma^{-1}-I)f(n)=f(n-1)-f(n).
\]
Where $\Gamma^{-1} f(n)=f(n-1)$, and the corresponding ``$\circ$"
operation is
\[\triangle^*\circ
f=\sum^{\infty}_{i=0}\binom{j}{i}(\triangle^{*i}f)(n+i-j)\triangle^{*j-i}.
\]
Then we obtain the adjoint ring $F(\triangle^*)$ to the
$F(\triangle)$, and the formal adjoint to $R\in F(\triangle)$ is
$R^*\in F(\triangle^*)$, defined by $R^*=\sum_{j=-\infty}^d
\triangle^{*j}\circ f_j(n)$. The $*$ operation satisfies $(F\circ
G)^*=G^*\circ F^*$ for two operators and $f(n)^*=f(n)$ for a
function.

         The discrete KP-hierarchy \cite{hi} is a family of evolution equation in
infinitely many variables $t=(t_1,t_2,\cdots)$
\begin{equation}
\frac{\partial L}{\partial t_i}=[(L^i)_+, L].
\end{equation}
Where $L$ is a general first-order pseudo difference operator
\begin{equation} \label{laxoperatordkp}
L=\triangle + \sum_{j=0}^{\infty} f_j(n)\triangle^{-j}.
\end{equation}
Similar to KP, $L$ also can be generate by dressing operator W
\[W(n;t)=1+\sum^\infty_{j=1}w_j(n;t)\triangle^{-j},
\]
and
\begin{equation}
L=W\circ\triangle\circ W^{-1}.\label{88}
\end{equation}
         There are wave function $w(n;t,z)$ and adjoint wave
function $w^*(n;t,z)$,
\begin{align}
w(n;t,z)&=W(n;t)(1+z)^n exp(\sum^\infty_{i=1}t_i z^i)\notag\\
        &=(1+\frac{w_1(n;t)}{z}+\frac{w_2(n;t)}{z^2}+\cdots)(1+z)^n exp(\sum^\infty_{i=1}t_i
        z^i)\label{41}
\end{align}
and
\begin{align}
w^*(n;t,z)&=(W^{-1}(n-1;t))^*(1+z)^{-n} exp(\sum^\infty_{i=1}-t_i
z^i)\notag\\
&=(1+\frac{w_1^*(n;t)}{z}+\frac{w_2^*(n;t)}{z^2}+\cdots)(1+z)^{-n}
exp(\sum^\infty_{i=1}-t_i z^i).
\end{align}
 Also a tau function $\tau_\triangle=\tau(n;t)$  for dKP
exists \cite{hi}, which satisfies that
\begin{equation}
w(n;t,z)=\frac{\tau(n;t-[z])}{\tau(n;t)}(1+z)^n
exp(\sum^\infty_{i=1}t_i z^i)\label{21}
\end{equation}
and
\begin{equation}
w^*(n;t,z)=\frac{\tau(n;t+[z])}{\tau(n;t)}(1+z)^{-n}
exp(\sum^\infty_{i=1}-t_i z^i),\label{22}
\end{equation}
where $[z]=(\frac{1}{z},\frac{1}{2z^2},\frac{1}{3z^3},\cdots)$.

By comparing (\ref{41}) with (\ref{21}) one  finds that  $w_i$ are
expressed by $\tau(n;t)$,and further obtains dressing operator
$W(n;t)$ as following form
\begin{equation}
W(n;t)=1-(\frac{1}{\tau(n;t)}\partial_{t_1}\tau(n;t))\triangle^{-1}+(\frac{1}{2\tau(n;t)}(\partial^2_{t_1}-\partial_{t_2})\tau(n;t))\triangle^{-2}+\cdots\label{89}.
\end{equation}
And taking it back into (\ref{88}), then comparing (\ref{88}) and (\ref{laxoperatordkp}), all of dynamical variables  $\{f_i(n)\}$ can
be expanded by $\tau(n;t)$. The first few of $\{f_i\}$ can be written out  explicitly below,
\begin{eqnarray}
f_0(n)&=&w_1(n)-w_1(n+1)=\triangle\partial_{t_1}\ln\tau(n;t),\\
f_{-1}(n)&=&w_1(n+1)w_1(n)-w_2(n+1)-w_1^2(n)+w_2(n)-(\triangle w_1)(n)\nonumber \\
&=&(\triangle w_1(n)) (w_1(n)-1)- \triangle w_2(n) \nonumber \\
&=&\big(\triangle \partial_{t_1}\ln\tau(n;t) \big) \big(\partial_{t_1}\ln\tau(n;t)+1 \big)-
\triangle\big(\frac{1}{2\tau(n;t)}(\partial^2_{t_1}-\partial_{t_2})\tau(n;t)\big) \\
&\vdots \nonumber  &
\end{eqnarray}
On the other hand, from the  Sato equation of the dKP hierarchy,  we can get an important formula on residue of $L^n$,i.e.
\begin{equation}\label{resLi}
\partial^2_{t_1t_i}\ln\tau(n;t)= resL^i, \quad  i\geq 1,
\end{equation}
which leads to another useful forms of $f_i$ through $\tau(n;t)$.For examples,
\begin{eqnarray}
f_{-1}(n)&=&\partial^2_{t_1}\ln\tau(n;t),\\
\partial^2_{t_1,t_2}\ln\tau(n;t)&=& \triangle
f_{-1}(n)+f_{-2}(n+1)+f_0(n)f_{-1}(n)+f_{-1}(n)f_0(n-1)+f_{-2}(n),\\
&\vdots\nonumber &
\end{eqnarray}
Note that,in general,  (\ref{resLi}) implies recursion relation of $f_{-i}(n)$ if $i\geq 2$, one can not get explicit
form of $f_{-i}$ form it  except $f_{-1}$. This is a crucial difference of dKP hierarchy from KP hierarchy.

Now we prove some useful properties for the operators which are used later. \\
{\sl {\bf  Lemma 2.1}}  For $f\in F$ and $\triangle$, $\Gamma$ as
above, the following identities hold.
\begin{align}
&(1)\quad \triangle\circ\Gamma=\Gamma\circ\triangle  \\
&(2)\quad \triangle^*=-\triangle\circ\Gamma^{-1}  \\
&(3)\quad
(\triangle^{-1})^*=(\triangle^*)^{-1}=-\Gamma\circ\triangle^{-1} \\
&(4)\quad f\circ\triangle^{-1}=\sum_{i\geq
0}\triangle^{-i-1}\circ\triangle^i(\Gamma f)\\
&(5)\quad \triangle^{-1}\circ
f\circ\triangle^{-1}=(\triangle^{-1}f)\circ\triangle^{-1}-\triangle^{-1}\circ
\Gamma(\triangle^{-1} f)\label{85}
\end{align}

{\sl {\bf  Proof:}}     For $\forall g(n)\in F$,\\
(1) \[ (\triangle\circ\Gamma)
g(n)=g(n+2)-g(n+1)=\Gamma(g(n+1)-g(n))=(\Gamma\circ\triangle) g(n)
\]
(2)
\[\triangle^*g(n)=-(g(n)-g(n-1))=(-\triangle\circ\Gamma^{-1})g(n)
\]
(3)
\[I=(\triangle^*)^{-1}\circ\triangle^*=(-\Gamma\circ\triangle^{-1})\circ(-\triangle\circ\Gamma^{-1})
\]
(4) \begin{align}\sum_{i\geq
0}\triangle^{-i-1}\circ\triangle^i(\Gamma f)&=\sum_{i\geq
0}\sum^{\infty}_{i=0}\binom{-i-1}{j}(\triangle^{i+j}
f)(n-i-j)\circ\triangle^{-i-j-1}(by\quad (\ref{81}))\notag\\
&=\sum^\infty_{l=0}\sum^l_{j=0}\binom{j-l-1}{j}(\triangle^l
f)(n-l)\circ\triangle^{-l-1}(let\quad l=i+j)\notag\\
&=f\circ\triangle^{-1}.\notag\\(Since\quad &when\quad l\geq1\quad
\sum^l_0\binom{j-l-1}{j}=\sum^l_0\binom{l}{j}(-1)^j=0\notag)\notag
\end{align}
(5)We define\footnote{This definition is coincide with (\ref{81})}
\begin{equation}
\triangle^{-1}=(\Gamma-I)^{-1}=\frac{\Gamma^{-1}}{I-\Gamma^{-1}}=\sum_{i=1}^{\infty}(\Gamma^{-1})^i.\label{13}
\end{equation}
And thus
\begin{align}
&(\sum_{i=1}^{\infty}f(n-i))(\sum_{j=1}^{\infty}g(n-j))  \notag \\
=&\sum_{i=1}^{\infty}\Gamma^{-i}(f(n)\sum_{m=1}^{\infty}g(n-m))+\sum_{j=1}^{\infty}\Gamma^{-j}(g(n)\sum_{m=1}^{\infty}f(n-m+1)).\label{82}
\end{align}
By using (\ref{13}), equation (\ref{82}) implies
\[((\triangle^{-1}f)\circ\triangle^{-1})g(n)=(\triangle^{-1}\circ
f\circ\triangle^{-1})g(n)+(\triangle^{-1}\circ \Gamma(\triangle^{-1}
f))g(n),
\]
which finishes the proof because $g(n)$ is an arbitrary
function.$\hfill{\Box}$

%%%%%%%%%%%%%%%%%%%%%%%%%%%%%%%%%%%%%%%%%%%%%%%%%%%%%%%%%

%\begin{center}
\section*{\S 3.  The determinant representation }
%\end{center}
\setcounter{section}{3}
\setcounter{equation}{0}

             In this section, we shall construct the determinant representation of the gauge transformations.
To this end, we start with two elementary  types of gauge transformation
operators of the dKP hierarchy\cite{oe}:
\begin{align}
&T_d(\phi)=\Gamma(\phi)\circ\triangle\circ\phi^{-1}=\triangle-\frac{\triangle\phi}{\phi},\label{31}\\
&T_i(\psi)=\Gamma^{-1}(\psi^{-1})\circ\triangle^{-1}\circ\psi=(\triangle+\frac{\triangle(\Gamma^{-1}(\psi))}{\psi})^{-1}.\label{32}
\end{align}
Where $\phi$ and $\psi$ are called eigenfunctions and adjoint
eigenfunctions of  $L$ for gauge transformation, which satisfy the
following dynamic equation of dKP system
\[\partial_{t_n}\phi=(L^n)_+\phi,
\]
\[\partial_{t_n}\psi=-(L^n_+)^*\psi.
\]
Here the definition of adjoint eigenfunction are different from
(\ref{22}).

Set $\{\phi_1,\phi_2,\cdots,\phi_n\}$ and $\{\psi_1,\psi_2,\cdots,\psi_n\}$ be two sets of $n$ different
eigenfunctions and  $n$ different adjoint eigenfunctions of the dKP hierarchy associated with $L$. Then under the
transformation (\ref{31}), the corresponding transformations of \{$L, \phi,\psi$ \} are
\begin{align}
&L\rightarrow L^{(1)}=T_d(\phi_1)\circ L\circ T_d(\phi_1)^{-1}, \nonumber \\
&\phi\rightarrow\phi^{(1)}=T_d(\phi_1)\phi=\Gamma\phi-(\Gamma\phi_1)\phi^{-1}_1\phi,\\
&\psi\rightarrow\psi^{(1)}=(T_d(\phi_1)^*)^{-1}\psi=\frac{\Gamma(\triangle^{-1}(\phi_1\psi))}{\Gamma\phi_1}.
\end{align}
Similarly under the transformation (\ref{32}), the corresponding transformations of \{$L, \phi,\psi$ \}
are
\begin{align}
&L\rightarrow L^{(1)}=T_i(\psi_1)\circ L\circ T_i(\psi_1)^{-1}, \nonumber \\
&\phi\rightarrow\phi^{(1)}=T_i(\psi_1)\phi=\frac{\triangle^{-1}(\phi\psi_1)}{\Gamma^{-1}(\psi_1)},\\
&\psi\rightarrow\psi^{(1)}=(T_i(\psi_1)^*)^{-1}\psi=-\psi\psi_1^{-1}\Gamma^{-1}(\psi_1)+\Gamma^{-1}(\psi).
\end{align}
Where $\phi^{(1)},\psi^{(1)}$ are corresponding eigenfunction and
adjoint eigenfunction to $L^{(1)}$. Note $T_d$ and $T_i$  will annihilate their generation function,i.e.,
\begin{equation}\label{annihilate}
T_d(\phi_1)\phi_1=0, \quad (T^{-1}_i(\psi_1))^*\psi_1=0,
\end{equation}
which is a crucial fact to find the representation of the $T_{n+k}$ below.

       Now we consider the successive gauge transformation including the two type transformations. In \cite{oe}, only n-fold $T_d$ or
n-fold binary DT was considered. However we shall study a general gauge transformation operator $T_{n+k}$ which consists of
n-fold $T_d$ and k-fold $T_i$ in the following. We define the operator as following,
\begin{equation}
T_{n+k}=T_i(\psi_k^{(n+k-1)})\circ\cdots\circ T_i(\psi_1^{(n)})\circ
T_d(\phi_n^{(n-1)})\circ\cdots\circ T_d(\phi_2^{(1)})\circ
T_d(\phi_1),
\end{equation}
in which
\[\phi_i^{(i-1)}=T_d(\phi_{i-1}^{(i-2)})\circ
T_d(\phi_{i-2}^{(i-3)})\cdots T_d(\phi_{2}^{(1)})\circ
T_d(\phi_1)\cdot\psi_i,
\]
\[\psi_j^{(n)}=(T_d(\phi_n^{(n-1)})^{-1})^*\cdots
(T_d(\phi_2^{(1)})^{-1})^*\circ (T_d(\phi_1)^{-1})^*\cdot\psi_j,
\]
\[\psi_j^{(n+l)}=(T_i(\psi_l^{(n+l-1)})^{-1})^*\cdots(T_i(\psi_2^{(n+1)})^{-1})^*\circ
(T_i(\psi_1^{(n)})^{-1})^*\cdot\psi_j^{(n)}.
\]
That means under transformation (3.7), we have
\begin{align}
&L\xrightarrow{T_d(\phi_1)}
L^{(1)}\xrightarrow{T_d(\phi_2^{(1)})}L^{(2)}\rightarrow\cdots\rightarrow
L^{(n-1)}\xrightarrow{T_d(\phi_n^{(n-1)})} L^{(n)}\nonumber\\
&\xrightarrow{T_i(\psi_1^{(n)})}
L^{(n+1)}\xrightarrow{T_i(\psi_2^{(n+1)})}
L^{(n+2)}\rightarrow\cdots\rightarrow
L^{(n+k-1)}\xrightarrow{T_i(\psi_k^{(n+k-1)})}L^{(n+k)}.
\end{align}
and $\{\phi_j^{(n)},\psi_j^{(n)}\}$ are corresponding eigenfunctions
and adjoint eigenfunctions of $L^{(n)}$.

           Now we show that the operator $T_{n+k}$ has a determinant representation, which
gives the explicit form of $T_{n+k}$ and is useful for direct
computation.
           First we introduce a discrete Wronskian determinant.
           Define

\begin{eqnarray}
W_n^\triangle(\phi_1,\cdots,\phi_n)=
  \begin{vmatrix}
  \phi_1                     & \phi_2                      &\cdots  & \phi_n                      \\
  \triangle\phi_1            & \triangle\phi_2             &\cdots  & \triangle\phi_n             \\
  \vdots                     & \vdots                      &\cdots  & \vdots                      \\
  \triangle^{n-k-1}\phi_1    & \triangle^{n-k-1}\phi_2     &\cdots  & \triangle^{n-k-1}\phi_n
  \end{vmatrix},
\end{eqnarray}
and more general

\begin{eqnarray}
IW_{n+k}^\triangle(\psi_k,\cdots,\psi_1;\phi_1,\cdots,\phi_n)=
  \begin{vmatrix}
  \triangle^{-1}\psi_k\phi_1 & \triangle^{-1}\psi_k\phi_2  &\cdots  & \triangle^{-1}\psi_k\phi_n  \\
  \vdots                     & \vdots                      &\cdots  & \vdots                      \\
  \triangle^{-1}\psi_1\phi_1 & \triangle^{-1}\psi_1\phi_2  &\cdots  & \triangle^{-1}\psi_1\phi_n  \\
  \phi_1                     & \phi_2                      &\cdots  & \phi_n                      \\
  \triangle\phi_1            & \triangle\phi_2             &\cdots  & \triangle\phi_n             \\
  \vdots                     & \vdots                      &\cdots  & \vdots                      \\
  \triangle^{n-k-1}\phi_1    & \triangle^{n-k-1}\phi_2     &\cdots  & \triangle^{n-k-1}\phi_n
  \end{vmatrix},
\end{eqnarray}

which shall be used in the following theorem.

 {\sl {\bf  Theorem 3.1}} If
$n>k$, $T_{n+k}$ and $T_{n+k}^{-1}$ has the following determinant
representation:

\begin{eqnarray}
  T_{n+k}=\frac{1}{IW_{n+k}^\triangle(\psi_k,\cdots,\psi_1;\phi_1,\cdots,\phi_n)}\cdot
  \begin{vmatrix}
  \triangle^{-1}\psi_k\phi_1 &\cdots  & \triangle^{-1}\psi_k\phi_n & \triangle^{-1}\circ\psi_k\\
  \vdots                     &\cdots  & \vdots                     & \vdots\\
  \triangle^{-1}\psi_1\phi_1 &\cdots  & \triangle^{-1}\psi_1\phi_n & \triangle^{-1}\circ\psi_1\\
  \phi_1                     &\cdots  & \phi_n                     & 1\\
  \triangle\phi_1            &\cdots  & \triangle\phi_n            & \triangle\\
  \vdots                     &\cdots  & \vdots                     & \vdots\\
  \triangle^{n-k}\phi_1      &\cdots  & \triangle^{n-k}\phi_n      & \triangle^{n-k}
  \end{vmatrix},\label{63}
\end{eqnarray}

\begin{eqnarray}
   T_{n+k}^{-1}=
   \begin{vmatrix}
   \phi_1\circ\triangle^{-1} &\Gamma(\triangle^{-1}\psi_k\phi_1)&\cdots&\Gamma(\triangle^{-1}\psi_1\phi_1)&\Gamma(\phi_1)& \cdots &\Gamma(\triangle^{n-k-2}\phi_1)\\
   \phi_2\circ\triangle^{-1} &\Gamma(\triangle^{-1}\psi_k\phi_2)&\cdots&\Gamma(\triangle^{-1}\psi_1\phi_2)&\Gamma(\phi_2)& \cdots &\Gamma(\triangle^{n-k-2}\phi_2)\\
   \vdots                    &\vdots                            &\cdots&\vdots                            &\vdots        & \cdots &\vdots\\
   \phi_n\circ\triangle^{-1} &\Gamma(\triangle^{-1}\psi_k\phi_n)&\cdots&\Gamma(\triangle^{-1}\psi_1\phi_n)&\Gamma(\phi_n)& \cdots &\Gamma(\triangle^{n-k-2}\phi_n)
   \end{vmatrix}\nonumber\\
   \cdot
   \frac{(-1)^{n-1}}{\Gamma(IW_{n+k}^{\triangle}(\psi_k,\cdots,\psi_1;\phi_1,\cdots,\phi_n)}.
   \label{610}
\end{eqnarray}
here the determinant of $T_{n+k}$ expand by the last column and the
functions are on the left hand with the action $"\circ"$. As for
$T_{n+k}^{-1}$ the determinant expand by the first column and the
functions are on the right hand with the action $"\circ"$ too.

{\sl {\bf  Proof:}}    For $n>k$, we know the highest order item of
$T_{n+k}$ is $\triangle^{n-k}$. So we can assume $T_{n+k}$ has the
following form
\begin{equation}
T_{n+k}=\sum_{i=0}^{n-k}a_i\circ\triangle^i+\sum_{i=-k}^{-1}a_i\circ\triangle^{-1}\circ\psi_{|i|},\quad
a_{n-k}=1, \label{62}
\end{equation}
\begin{equation}
T_{n+k}^{-1}=\sum_{j=1}^n\phi_j\circ\triangle^{-1}\circ b_j.
\label{64}
\end{equation}
It is easy to find
\begin{align}
&T_{n+k}\cdot\phi_i=0, \quad i=1,2,\cdots,n ,\label{66}\\
&(T_{n+k}^{-1})^*\cdot\psi_j=0, \quad j=1,2,\cdots,k ,\label{65}
\end{align}
from (\ref{annihilate}),
which give the following algebraic system on $a_i$ and $b_j$,
\begin{align}
a_{-k}\triangle^{-1}(\psi_k\phi_i)+\cdots+a_{-1}\triangle^{-1}(\psi_1\phi_i)+a_0\phi_i+&\cdots+a_{n-k-1}\triangle^{n-k-1}\phi_i=-\triangle^{n-k}\phi_i \label{61}\\
&i=1,2,\cdots,n,\nonumber
\end{align}
\begin{align}
b_1\Gamma(\triangle^{-1}(\psi_j\phi_1))+b_2\Gamma(\triangle^{-1}(\psi_j\phi_2))&+\cdots+b_n\Gamma(\triangle^{-1}\psi_j\phi_n))=0 \label{69}\\
&j=1,2,\cdots,k.\nonumber
\end{align}
Then with the help of (\ref{62}), by Gramer  rule the solution of
(\ref{61}) is given by
\begin{eqnarray}
   a_{-i}=\frac{-1}{IW_{n+k}^{\triangle}(\psi_k,\cdots,\psi_1;\phi_1,\cdots,\phi_n)} \cdot \nonumber\hspace{8cm}\\
   \begin{vmatrix}
   \triangle^{-1}\psi_k\phi_1 &\cdots &\triangle^{-1}\psi_{i+1}\phi_1 &\triangle^{n-k}\phi_1 &\triangle^{-1}\psi_{i-1}\phi_1 &\cdots &\triangle^{-1}\psi_1\phi_1 &\phi_1 &\cdots &\triangle^{n-k-1}\phi_1\\
   \vdots   &\cdots &\vdots&\vdots&\vdots&\cdots &\vdots&\vdots&\cdots &\vdots\\
   \triangle^{-1}\psi_k\phi_n &\cdots &\triangle^{-1}\psi_{i+1}\phi_n &\triangle^{n-k}\phi_n &\triangle^{-1}\psi_{i-1}\phi_n &\cdots &\triangle^{-1}\psi_1\phi_n &\phi_n &\cdots &\triangle^{n-k-1}\phi_n
  \end{vmatrix}, \nonumber
 \\
\end{eqnarray}
\begin{eqnarray}
   a_i=\frac{-1}{IW_{n+k}^{\triangle}(\psi_k,\cdots,\psi_1;\phi_1,\cdots,\phi_n)} \times \nonumber\hspace{8cm}\\
   \begin{vmatrix}
   \triangle^{-1}\psi_k\phi_1 &\cdots &\triangle^{-1}\psi_1\phi_1 &\phi_1 &\cdots &\triangle^{i-1}\phi_1 &\triangle^{n-k}\phi_1 &\triangle^{i+1}\phi_1 &\cdots &\triangle^{n-k-1}\phi_1\\
   \vdots   &\cdots &\vdots&\vdots&\cdots&\vdots &\vdots&\vdots&\cdots &\vdots\\
   \triangle^{-1}\psi_k\phi_n &\cdots &\triangle^{-1}\psi_1\phi_n &\phi_n &\cdots &\triangle^{i-1}\phi_n &\triangle^{n-k}\phi_n &\triangle^{i+1}\phi_n &\cdots &\triangle^{n-k-1}\phi_n
  \end{vmatrix}.
\end{eqnarray}
So we obtain the representation of $T_{n+k}$ as (\ref{63}).

          Next we prove the determinant representation of $T_{n+k}^{-1}$.
Because $T_{n+k}\circ T_{n+k}^{-1}=1$, For the rationality of the
definition of $T_{n+k}^{-1}$ in (\ref{64}) we have to verify
\[(T_{n+k}\circ\ T_{n+k}^{-1})_-=I_-=0.
\]
Taking (\ref{62}) and (\ref{64}) into the left hand side ,then
\begin{equation}
(T_{n+k}\circ T_{n+k}^{-1})_-=\sum_{i=0}^{n-k}\sum_{j=1}^n
a_i(\triangle^i\phi_j)\circ\triangle^{-1}\circ
b_j+\sum_{i=-k}^{-1}\sum_{j=1}^n
a_i\circ\triangle^{-1}\circ\psi_{|i|}\phi_j\circ\triangle^{-1}\circ
b_j.\label{84}
\end{equation}
Furthermore, we can rewrite the second term in (\ref{84}) from
(\ref{85})
\begin{eqnarray}
    (T_{n+k}\circ T_{n+k}^{-1})_-&=&\sum_{i=0}^{n-k}\sum_{j=0}^n a_i(\triangle^i\phi_j)\circ\triangle^{-1}\circ b_j \nonumber\\
    &+&\sum_{j=0}^n a_{-k}\circ(\triangle^{-1}\psi_k\phi_j)\circ\triangle^{-1} \circ b_j-\sum_{j=0}^n a_{-k}\circ\triangle^{-1}\circ\Gamma(\triangle^{-1}\psi_k\phi_j) b_j \nonumber\\
    &+&\cdots \nonumber \\
    &+& \sum_{j=0}^n a_{-1}\circ(\triangle^{-1}\psi_1\phi_j)\circ\triangle^{-1} \circ b_j-\sum_{j=0}^n a_{-1}\circ\triangle^{-1}\circ\Gamma(\triangle^{-1}\psi_1\phi_j)
    b_j.\nonumber\\ \label{67}
\end{eqnarray}
Rearranging the right hand side, and using with (\ref{61}) and (\ref{69}), then (\ref{67}) becomes
\begin{eqnarray}
    (T_{n+k}\circ T_{n+k}^{-1})_-&=& \sum\limits_{j=1}^n T_{n+k}(\phi_j)\circ \triangle^{-1} \circ b_j\\
    & &- \sum\limits_{i=1}^k  a_i  \circ \triangle^{-1}  \circ  (T^{-1}_{n+k})^*(\psi_i) \\
    &=0,&
\end{eqnarray}
due to (\ref{66})  and (\ref{65}). On the other hand, by lemma 2.1, $T_{n+k}^{-1}$ has the following form
\begin{eqnarray}
    T_{n+k}^{-1}&=&\sum_{i=0}^{n-k-2}\sum_{j=1}^n\triangle^{-1-i}\circ \Gamma(\triangle^i\phi_j) b_j \nonumber \\
    &+&\triangle^{-n+k}\circ\sum_{j=0}^n\Gamma(\triangle^{n-k-1}\phi_j)b_j+\sum_{i=n-k}^{\infty}\sum_{j=0}^n\triangle^{-1-i}\circ\Gamma(\triangle^i\phi_j)b_j,
    \label{68}
\end{eqnarray}
and  the $T_{n+k}^{-1}$ may be expressed as
\begin{equation}
T_{n+k}^{-1}=\triangle^{-n+k}+(lower \quad oder \quad
item),\label{86}
\end{equation}
according to the $T_{n+k}\circ T_{n+k}^{-1}=I$ and the form of
$T_{n+k}$. Then we get the following algebraic equations on $b_i$,
\begin{eqnarray}
  \begin{cases}
  \Gamma(\triangle^{-1}(\psi_k\phi_1))b_1+\Gamma(\triangle^{-1}(\psi_k\phi_2))b_2+\cdots+\Gamma(\triangle^{-1}\psi_k\phi_n))b_n=0\\
  \hspace{3cm}\vdots\\
  \Gamma(\triangle^{-1}(\psi_1\phi_1))b_1+\Gamma(\triangle^{-1}(\psi_1\phi_2))b_2+\cdots+\Gamma(\triangle^{-1}\psi_1\phi_n))b_n=0\\
  \Gamma(\phi_1)b_1+\Gamma(\phi_2)b_2+\cdots+\Gamma(\phi_n)b_n=0\\
  \Gamma(\triangle\phi_1)b_1+\Gamma(\triangle\phi_2)b_2+\cdots+\Gamma(\triangle\phi_n)b_n=0\\
  \hspace{3cm}\vdots\\
  \Gamma(\triangle^{n-k-2}\phi_1)b_1+\Gamma(\triangle^{n-k-2}\phi_2)b_2+\cdots+\Gamma(\triangle^{n-k-2}\phi_n)b_n=0\\
  \Gamma(\triangle^{n-k-1}\phi_1)b_1+\Gamma(\triangle^{n-k-1}\phi_2)b_2+\cdots+\Gamma(\triangle^{n-k-1}\phi_n)b_n=1
  \end{cases}.\label{87}
\end{eqnarray}
Note the first $k$ equations are given by (\ref{69}) and the last
$n-k$ equations are given by comparing the corresponding terms in
(\ref{68}) with (\ref{86}). Solving (\ref{87}) gives
\begin{eqnarray}
    b_i=\frac{1}{\Gamma(IW_{n+k}^{\triangle}(\psi_k,\cdots,\psi_1;\phi_1,\cdots,\phi_n))} \cdot \hspace{7cm}\nonumber\\
    \begin{vmatrix}
    \Gamma(\triangle^{-1}\psi_k\phi_1)   &\cdots &\Gamma(\triangle^{-1}\psi_k\phi_{i-1}) &0        &\Gamma(\triangle^{-1}\psi_k\phi_{i+1}) &\cdots &\Gamma(\triangle^{-1}\psi_k\phi_n)\\
    \vdots                               &\cdots &\vdots                                 &\vdots   &\vdots                                 &\cdots &\vdots                            \\
    \Gamma(\triangle^{-1}\psi_1\phi_1)   &\cdots &\Gamma(\triangle^{-1}\psi_1\phi_{i-1}) &0        &\Gamma(\triangle^{-1}\psi_1\phi_{i+1}) &\cdots &\Gamma(\triangle^{-1}\psi_1\phi_n)\\
    \Gamma(\phi_1)                       &\cdots &\Gamma(\phi_{i-1})                     &0        &\Gamma(\phi_{i+1})                     &\cdots &\Gamma(\phi_n)                    \\
    \vdots                               &\cdots &\vdots                                 &\vdots   &\vdots                                 &\cdots &\vdots                            \\
    \Gamma(\triangle^{n-k-1}\phi_1)      &\cdots &\Gamma(\triangle^{n-k-1}\phi_{i-1})    &1        &\Gamma(\triangle^{n-k-1}\phi_{i+1})    &\cdots &\Gamma(\triangle^{n-k-1}\phi_n)
    \end{vmatrix},
\end{eqnarray}
and take it back to (\ref{64}), we obtain (\ref{610}). So we finish
the proof.$\hfill{\Box}$

We notice that when use $\partial$ instead of $\triangle$, and set $n=0,f_0=0$, all results will approach to the case of KP automatically\cite{hlc}.For the case of $n=k$, we have the following

{\sl {\bf  Theorem 3.2}}      When $n=k$,
\begin{eqnarray}
  T_{n+n}=\frac{1}{IW_{n+n}^{\triangle}(\psi_n,\cdots,\psi_1;\phi_1,\cdots,\phi_n)}\cdot
  \begin{vmatrix}
  \triangle^{-1}\psi_n\phi_1 &\cdots  & \triangle^{-1}\psi_n\phi_n & \triangle^{-1}\circ\psi_n\\
  \vdots                     &\cdots  & \vdots                     & \vdots\\
  \triangle^{-1}\psi_1\phi_1 &\cdots  & \triangle^{-1}\psi_1\phi_n & \triangle^{-1}\circ\psi_1\\
  \phi_1                     &\cdots  & \phi_n                     & 1
  \end{vmatrix},
\end{eqnarray}
\begin{eqnarray}
   T_{n+n}^{-1}=
   \begin{vmatrix}
   -1                        &\psi_n                             &\cdots &\psi_1  \\
   \phi_1\circ\triangle^{-1} &\Gamma(\triangle^{-1}\psi_n\phi_1) &\cdots &\Gamma(\triangle^{-1}\psi_1\phi_1)\\
   \vdots                    &\vdots                             &\cdots &\vdots                 \\
   \phi_n\circ\triangle^{-1} &\Gamma(\triangle^{-1}\psi_n\phi_n) &\cdots &\Gamma(\triangle^{-1}\psi_1\phi_n)
   \end{vmatrix}
\cdot\frac{(-1)}{\Gamma(IW_{n+n}^{\triangle}(\psi_n,\cdots,\psi_1;\phi_1,\cdots,\phi_n)}.
\end{eqnarray}

{\sl {\bf  Proof:}}    Assume
\[T_{n+n}=1+ \sum_{i=-n}^{-1}a_i\circ\triangle^{-1}\circ\psi_{|i|},
\]
\[T_{n+n}^{-1}=1+ \sum_{j=1}^n\phi_j\circ\triangle^{-1}\circ b_j.
\]
The remaining proof is as Theorem 3.1, so we omit it.

Now we obtain the determinant representation of $(T_{n+k}^{-1})^*$
by a direct computation, which is useful in the following context.

{\sl   {\bf   Corollary 3.3}} When $n>k$,
\begin{eqnarray}
    (T_{n+k}^{-1})^*=\frac{-1}{\Gamma(IW_{n+k}^{\triangle}(\psi_k,\cdots,\psi_1;\phi_1,\cdots,\phi_n))}\cdot\hspace{7cm}\nonumber\\
    \begin{vmatrix}
    \Gamma(\triangle^{-1}\psi_k\phi_1) &\cdots &\Gamma(\triangle^{-1}\psi_1\phi_1) &\Gamma(\phi_1) &\cdots &\Gamma(\triangle^{n-k-2}\phi_1) &\Gamma\circ\triangle^{-1}\circ\phi_1\\
    \vdots                            &\cdots &\vdots                            &\vdots         &\cdots &\vdots                              &\vdots\\
    \Gamma(\triangle^{-1}\psi_k\phi_n) &\cdots &\Gamma(\triangle^{-1}\psi_1\phi_n) &\Gamma(\phi_n) &\cdots &\Gamma(\triangle^{n-k-2}\phi_n) &\Gamma\circ\triangle^{-1}\circ\phi_n
    \end{vmatrix}.\nonumber
\end{eqnarray}

\section*{\S 4 The transformed dKP hierarchy }
\setcounter{section}{4}
\setcounter{equation}{0}
 We are now in a position to give the transformed KP by using results in last
 section. In other hand we find the $\tau_\triangle^{(n+k)}$ function under the transformation of
 $T_{n+k}$. The existence of $\tau_\triangle$ function for dKP is proved in \cite{hi}.

{\sl {\bf Lemma 4.1}} Under the $T_d(\phi_1)$, we have
\begin{equation}
  \tau_\triangle\rightarrow\tau^{(1)}_\triangle=\phi_1\tau_\triangle,
\end{equation}
and under $T_i(\psi_1)$,
\begin{equation}
  \tau_\triangle\rightarrow\tau^{(1)}_\triangle=\psi_1(n-1)\tau_\triangle\label{52}.
\end{equation}

{\sl {\bf Proof}}:\quad\quad Taking an initial Lax operator $L$, we
get a transformed one
\[L^{(1)}=\triangle+f_0^{(1)}+f_{-1}^{(1)}\triangle^{-1}+\cdots.
\]
By directly computation
\[L^{(1)}=T_d(\phi_1)\circ L \circ
T_d(\phi_1)^{-1}=\triangle+\triangle(\frac{\triangle\phi_1}{\phi_1})+f_0(n+1)+\dots
\]
So
\[f_0^{(1)}=\triangle(\frac{\triangle\phi_1}{\phi_1})+f_0(n+1).
\]
This implies
\begin{equation}
f_0^{(1)}=\triangle\partial_{t1}\ln\phi_1+f_0, \label{51}
\end{equation}
by using the equation
\[(\phi_1)_{t1}=L_+\phi_1=(\triangle+f_0)\phi_1.
\]
Moreover, take $f_0=\triangle(\partial_{t_1}\ln\tau_\triangle)$ back
into the above equation, then

\[f_0^{(1)}=\triangle\partial_{t1}\ln\phi_1+\triangle\partial_{t1}\ln\tau_\triangle=\triangle\partial_{t1}\ln\phi_1\tau_\triangle.
\]
and this shows
\[\tau_\triangle^{(1)}=\phi_1\tau_\triangle,
\]
because $f_0^{(1)}=\triangle\partial_{t_1}\ln\tau_\triangle^{(1)}$. The proof of (\ref{52}) is analogous to the above. So it is
omitted.$\hfill{\Box}$

  By a direct computation with the help  of  above theorem, we have the following theorem.

{\sl {\bf Theorem 4.2}} When $k=0$ under the transformation of
$T_{n+0}=T_n$, there are
\begin{equation}\label{phin}
\phi^{(n)}=T_n\cdot\phi=\frac{W^\triangle_{n+1}(\phi_1,,\cdots,\phi_n,\phi)}{W^\triangle_n(\phi_1,\cdots,\phi_n)},
\end{equation}
\begin{equation}\label{psin}
  \psi^{(n)}=(T_n^{-1})^*\cdot\psi=
  \frac{(-1)^n\Gamma(IW^\triangle_{n+1}(\psi,\phi_1,\cdots,\phi_n))}{\Gamma(W^\triangle_n(\phi_1,\cdots,\phi_n))},
\end{equation}
and
\begin{equation}
  \tau_\triangle^{(n)}=W_n^\triangle(\phi_1,\cdots,\phi_n)\cdot\tau_\triangle.
\end{equation}
{\sl {\bf Proof:}}
The determinant representation of $T_n$ is given by $T_{n+k}|_{k=0}$, then substituting it into $\phi^{(n)}=T_n\phi$
 and $\psi^{(n)}=(T_n^{-1})^*\psi$, we can get results of this theorem. Furthermore, $\tau^{(n)}_\triangle$ is given by applying  the gauge
transformation repeatedly, i.e.,
\begin{eqnarray}
\tau_\triangle^{(n)}&&=\phi_n^{(n-1)}\tau_\triangle^{(n-1)}=\phi_n^{(n-1)}\phi_{n-1}^{(n-2)}\tau_\triangle^{(n-2)}=\phi_n^{(n-1)}\phi_{n-1}^{(n-2)}\cdots\phi_2^{(1)}\phi_1\tau_\triangle\notag\\
                      &&=\frac{W_n^\triangle(\phi_1,\cdots,\phi_n)}{W_{n-1}^\triangle(\phi_1,\cdots,\phi_{n-1})}\cdot\frac{W_{n-1}^\triangle(\phi_1,\cdots,\phi_{n-1})}{W_{n-2}^\triangle(\phi_1,\cdots,\phi_{n-2})}\cdots\frac{W_2^\triangle(\phi_1,\phi_2)}{W_1^\triangle(\phi_1)}\cdot\phi_1\cdot\tau_\triangle\notag\\
                      &&=W_n^\triangle(\phi_1,\cdots,\phi_n)\tau_\triangle.
\end{eqnarray}
$\hfill{\Box}$

{\sl {\bf  Theorem 4.3}} When $n>k$, under the $T_{n+k}$,
\begin{equation}
\phi^{(n+k)}=T_{n+k}\cdot\phi=\frac{IW^\triangle_{(n+1)+k}(\psi_k,\cdots,\psi_1;\phi_1,\cdots,\phi_n,\phi)}{IW^\triangle_{n+k}(\psi_k,\cdots,\psi_1;\phi_n,\cdots,\phi_1)},
\end{equation}
\begin{eqnarray}
  \psi^{(n+k)}=(T^{-1}_{n+k})^*\psi=\frac{(-1)^n\Gamma(IW_{n+(k+1)}^{\triangle}(\psi,\psi_k,\cdots,\psi_1;\phi_1,\cdots,\phi_n))}
   {\Gamma(IW_{n+k}^{\triangle}(\psi_k,\cdots,\psi_1;\phi_1,\cdots,\phi_n))}\\
   (\psi\ne\psi_i,i=1,\cdots,k),\notag
\end{eqnarray}
and
\begin{equation}
\tau_\triangle^{n+k}=(-1)^{kn}IW^\triangle_{k,n}(\psi_k,\cdots,\psi_1;\phi_1,\cdots,\phi_n)\cdot\tau_\triangle.
\end{equation}
{\sl {\bf Proof:}} The $\phi^{(n+k)}$ and $\psi^{(n+k)}$ are obtained by a direct application of determinant of $T_{n+k}$
and $(T^{-1}_{n+k})^*$. Moreover, $\tau^{(n+k)}_\triangle$ can be derived by following way,
\begin{eqnarray}
\tau_\triangle^{(n+k)}&&=\Gamma^{-1}(\psi_k^{(n+k-1)})\tau_\triangle^{(n+k-1)}=\Gamma^{-1}(\psi_k^{(n+k-1)})\Gamma^{-1}(\psi_{k-1}^{(n+k-2)})\tau_\triangle^{(n+k-2)}\notag\\
                      &&=\psi_k^{(n+k-1)}(n-1)\psi_{k-1}^{(n+k-2)}(n-1)\cdots\psi_2^{(n+1)}(n-1)\psi_1^{(n)}(n-1)\tau_\triangle^{(n)}\notag\\
                      &&=(-1)^n\frac{IW_{n+k}^\triangle(\psi_k,\cdots,\psi_1;\phi_1,\cdots,\phi_n)}{IW_{n+(k-1)}^\triangle(\psi_{k-1},\cdots,\psi_1;\phi_1,\cdots,\phi_n)}\cdot\notag\\
                      &&(-1)^n\frac{IW_{n+(k-1)}^\triangle(\psi_{k-1},\cdots,\psi_1;\phi_1,\cdots,\phi_n)}{IW_{n+(k-2)}^\triangle(\psi_{k-2},\cdots,\psi_1;\phi_1,\cdots,\phi_n)}\cdots\notag\\
                      &&\cdots\cdot(-1)^n\frac{IW^\triangle_{n+1}(\psi_1;\phi_1,\cdots,\phi_n)}{W_n^\triangle(\phi_1,\cdots,\phi_n)}W_n^\triangle(\phi_1,\cdots,\phi_n)\tau_\triangle\notag\\
                      &&=(-1)^{nk}IW^\triangle_{n+k}(\psi_k,\cdots,\psi_1;\phi_1,\cdots,\phi_n)\tau_\triangle.
\end{eqnarray} $\hfill{\Box}$

\section*{\S 5  Conclusions and discussions   }

We have established determinant representant of the gauge transformation operator $T_{n+k}$ of the dKP hierarchy, and further
given the transformed $\tau^{(n+k)}_\triangle$ from known  $\tau_\triangle$. The advantage of this representation  is compact
and systematic by comparing with an iterative way of gauge transformation. Also it is convenient for find solution of dKP
hierarchy. To illustrate our approach, we just have given some results on a special  chain of the gauge transformations.
Of course, it is possible to consider other complicated chain if it is needed. We notice that under the map: $\triangle\rightarrow
\partial$, $n\equiv0$, $f_0=0$, the whole dKP hierarchy approaches to the KP hierarchy, and our results in previous sections
can go back to the case of KP  hierarchy too.

Because of the differences between operators $\triangle$ and $\partial$ in dKP and KP hierarchy respectively, and the existence
of $\{f_0\}$, we know it is not a easy task to get simple closed  form of the dKP equation by $f_{-1}$, and then
to get explicit solution of dKP equation by gauge transformation as we have done for the KP equation. However, we can get
an explicit  solution of $f_{-1}$ through $\tau^{(n+k)}_{\triangle}$ starting from a known $\tau_{\triangle}=1$, which can be regarded as discrete
analogue of $u_1$ in KP hierarchy with a Lax operator $L=\partial +u_1\partial^{-1}+ u_2\partial^{-2}+\cdots$, and then to show the
difference between $f_{-1}$ and $u_1$. This is a way for us to understand the discrete effect of the KP hierarchy.
On the other hand, it is possible to find the additional symmetry and its algebraic structures of the dKP hierarchy due to the existence of
the tau function $\tau_{\triangle}=\tau{(n;t)}$\cite{hi}. We shall try to do it in a near future.

{\bf Acknowledgement}\\
{\small This work is supported partly by the NSFC grant of China
under No.10671187. He is also supported by the Program for NCET
under Grant No.NCET-08-0515. We thank Professor Li Yishen(USTC,
China) for long-term encouragements and supports. }

%%%%%%%%%%%%%%%%%%%%%%%%%%%%%%%%%%%%%%%%%%%%%%%%%%%%%%%%%%%%%%%%%%%%%%%%%%%%%%%%%%%%%%%%%%%%%%%%%%%%%%%%%%%%%%%%%%
%%%%%%%%%%%%%%%%%%%%%%%%%%%%%%%%%%%%%%%%%%%%%%%%%%%%%%%%%%%%%%%%%%%%%%%%%%%%%%%%%%%%%%%%%%%%%%%%%%%%%%%%%%%%%%%%%%%

\end{document}